\documentclass[11pt,a4paper]{article}

\usepackage{jheppub}

\usepackage[utf8]{inputenc}
\usepackage{graphicx}
\usepackage{epsfig}
\usepackage{subfigure}
\usepackage{color}
\usepackage{comment}
\usepackage{slashed}

\newcommand{\be}{\begin{equation}} 
\newcommand{\ee}{\end{equation}}
\newcommand{\bea}{\begin{eqnarray}} 
\newcommand{\eea}{\end{eqnarray}}

\newcommand{\bmp}{\noindent\begin{minipage}{16cm}}
\newcommand{\emp}{\end{minipage}\vskip 7mm} 
\def\lsim{\mathrel{\raise.3ex\hbox{$<$\kern-.75em\lower1ex\hbox{$\sim$}}}}
\def\gsim{\mathrel{\raise.3ex\hbox{$>$\kern-.75em\lower1ex\hbox{$\sim$}}}}

\newcommand{\intron}[1]{}

\title{Scale-invariant scalar field dark matter through the Higgs portal}

\author[a]{Catarina Cosme,}
\author[b]{Jo\~{a}o G.~Rosa}
\author[a]{and O. Bertolami}

\affiliation[a]{Departamento de F\'{\i}sica e Astronomia, Faculdade de Ci\^encias da Universidade do Porto and Centro de F\'{\i}sica do Porto, Rua do Campo Alegre 687, 4169-007, Porto, Portugal}
\affiliation[b]{Departamento de F\'{\i}sica da Universidade de Aveiro and CIDMA,  Campus de Santiago, 3810-183 Aveiro, Portugal}

\emailAdd{catarinacosme@fc.up.pt}     
\emailAdd{joao.rosa@ua.pt}
\emailAdd{orfeu.bertolami@fc.up.pt}

\abstract{We discuss the dynamics and phenomenology of an oscillating scalar field coupled to the Higgs boson that accounts for the dark matter in the Universe. The model assumes an underlying scale invariance such that the scalar field only acquires mass after the electroweak phase transition, behaving as dark radiation before the latter takes place. While for a positive coupling to the Higgs field the dark scalar is stable, for a negative coupling it acquires a vacuum expectation value after the electroweak phase transition and may decay into photon pairs, albeit with a mean lifetime much larger than the age of the Universe. We explore possible astrophysical and laboratory signatures of such a dark matter candidate in both cases, including annihilation and decay into photons, Higgs decay, photon-dark scalar oscillations and induced oscillations of fundamental constants. We find that dark matter within this scenario will be generically difficult to detect in the near future, except for the promising case of a 7 keV dark scalar decaying into photons, which naturally explains the observed galactic and extra-galactic 3.5 keV X-ray line. }

\keywords{Dark matter, Scalar Field, Higgs boson}
\arxivnumber{1802.09434}

\begin{document}
\maketitle


\section{Introduction}
\label{model}

Dark matter is one of the most important open puzzles of modern cosmology and fundamental physics. Although a particle explanation seems to be favoured by observational data, as opposed to e.g.~modified gravity theories, such putative new particles have so far evaded detection, with a wide range of masses and couplings to the Standard Model fields being still allowed.   

Weakly Interacting Massive Particles (WIMPs) are certainly the most popular dark matter candidates in the literature, corresponding to particles with masses typically within the GeV-TeV range that attained thermal equilibrium with the cosmic plasma in the early Universe and later decoupled to yield a frozen-out abundance. The so-called ``WIMP miracle", where the relic WIMP abundance matches the present dark matter abundance for weak-scale cross sections, makes such scenarios quite appealing, with a plethora of candidates within extensions of the Standard Model at the TeV scale. However, the lack of experimental evidence for such WIMPs and, in particular, the absence of novel particles at the LHC, strongly motivates looking for alternative scenarios.  

An interesting candidate for dark matter is a dynamical homogeneous scalar field that is oscillating about the minimum of its (quadratic) potential, and which can be seen as a condensate of low-momentum particles acting coherently as non-relativistic matter. Scalar fields are ubiquitous in extensions of the Standard Model including, for instance, the QCD axion in the Peccei-Quinn scenario to address the strong CP problem, supersymmetric theories and theories with extra compact spatial dimensions. Establishing the form of the interactions between such scalars and the Standard Model particles is of the utmost importance to detect dark matter either directly or indirectly, and an obvious possibility is the Higgs portal, where dark matter only interacts directly with the Higgs field, $\mathcal{H}$. A coupling of the form $g^2|\Phi|^2|\mathcal{H}|^2$ should generically appear for any complex or real scalar field, since it is not forbidden by any symmetries, except for the QCD axion and analogous pseudo-scalars where such an interaction is forbidden by a shift symmetry.

The Higgs portal for dark matter has been thoroughly explored in the context of scalar WIMP-like candidates \cite{Patt:2006fw,Bento:2000ah,Bento:2001yk,MarchRussell:2008yu,Biswas:2011td,Pospelov:2011yp,Mahajan:2012nc,Cline:2013gha,Enqvist:2014zqa,Kouvaris:2014uoa,Costa:2014qga,Duerr:2015aka,Han:2015dua,Han:2015hda}, but only a few proposals in the literature discuss the case of an oscillating scalar condensate \cite{Nurmi:2015ema,Tenkanen:2015nxa,Kainulainen:2016vzv,Bertolami:2016ywc,Cosme:2017cxk}. In this work, we aim to fill in this gap and consider a generic model for scalar field dark matter where, like all other known particles, the dark scalar acquires mass exclusively through the Higgs mechanism, i.e.~no bare scalar mass term in the Lagrangian is introduced for dark matter. While the Standard Model gauge symmetries forbid bare masses for chiral fermions and gauge bosons, this is not so for scalars, since $|\Phi|^2$ is always a gauge-invariant operator. Scalar mass terms are, however, forbidden if the theory is scale-invariant (or exhibits a conformal invariance). This has arisen some interest in the recent literature, with the possibility of dynamically generating both the Planck scale and the electroweak scale through a spontaneous breaking of scale-invariance. In fact, with the inclusion of non-minimal couplings to gravity allowed by scale-invariance, one can generate large hierarchies between mass scales from hierarchies between dimensionless couplings and naturally obtain an inflationary period in the early Universe, as shown in Refs. \cite{GarciaBellido:2011de, Bezrukov:2012hx, Ferreira:2016vsc, Ferreira:2016wem}.For other scenarios with scale-invariance and viable dark matter candidates, see also Refs. \cite{Heikinheimo:2013fta,Gabrielli:2013hma,Salvio:2014soa, Heikinheimo:2017ofk, Eichhorn:2017als, Salvio:2017xul}.

In this work we will pursue this possibility, considering a model of scalar-field dark matter with scale-invariant Higgs-portal interactions. We assume that scale invariance is spontaneously broken by some unspecified mechanism that generates both the Planck scale, $M_P$, and a negative squared mass for the Higgs field at the electroweak scale, thus working within an effective field theory where these mass scales are non-dynamical. The important assumption is that scale-invariance is preserved in the dark matter sector, such that the dark scalar only acquires mass after the electroweak phase transition (EWPT). In addition, our scenario has also a U(1) symmetry (or $\mathbb{Z}_2$ symmetry for a real scalar, as we discuss later on) that, if unbroken, ensures the stability of the dark scalar. Its dynamics are thus fully determined by its interaction with the Higgs boson and gravity, as well as its self-interactions, all parametrized by dimensionless couplings. The relevant interaction Lagrangian density is thus given by:
\begin{equation}
\mathcal{-L}_{int}=\pm\, g^{2}\left|\Phi\right|^{2}\left|\mathcal{H}\right|^{2}+\lambda_{\phi}\left|\Phi\right|^{4}+V\left(\mathcal{H}\right)+\xi R\left|\Phi\right|^{2}~,
\label{Lagrangian}
\end{equation}
where the Higgs potential, $V\left(\mathcal{H}\right)$, has the usual ``mexican hat" shape, $g$ is the coupling between the Higgs and the dark scalar and $\lambda_{\phi}$ is the dark scalar's self-coupling. The last term in Eq. (\ref{Lagrangian}) corresponds to a non-minimal coupling of the dark matter field to curvature, where $R$ is the Ricci scalar and $\xi$ is a constant. Note that such a Lagrangian density is an extension of the model that we considered in Ref. \cite{Bertolami:2016ywc} where self-interactions played no role in the dynamics, giving origin to a very light dark scalar, $m_{\phi}\sim\mathcal{O}\left(10^{-5}\,\mathrm{eV}\right)$, and therefore a very small coupling to the Higgs boson, $g\sim 10^{-16}$. Such feeble interactions make such a dark matter candidate nearly impossible to detect in the near future, and in this work we will show that the inclusion of self-interactions allows for heavier and hence more easily detectable dark scalars.

On the one hand, if the Higgs-dark scalar interaction has a positive sign, the U(1) symmetry remains unbroken in the vacuum and the dark scalar is stable. On the other hand, if the interaction has a negative sign, the U(1) symmetry may be spontaneously broken, which can lead to interesting astrophysical signatures, as we first observed in Ref. \cite{Cosme:2017cxk}. In this work we wish to provide a thourough discussion of the dynamics and phenomenology in both cases, highlighting their differences and similarities and exploring the potential to probe such a dark matter both in the laboratory and through astrophysical observations.

This work is organized as follows. In the next section we explore the scalar field dynamics from the inflationary period to the EWPT, where the Higgs-portal coupling plays a negligible role. In section \ref{after EWSB} we discuss the evolution of the scalar field after the EWPT, analyzing separately the cases where the Higgs-portal coupling is positive or negative, and computing the present dark matter abundance in both scenarios. We explore the phenomenology of these scenarios in section \ref{Pheno}, discussing possible astrophysical signatures and experiments that could test these scenarios in the laboratory. We summarize our conclusions and discuss possible avenues for future work in this subject in section \ref{conclusions}.


\section{Dynamics before Electroweak symmetry breaking}
\label{Dynamics before EWSB}
Before the EWPT, the dynamics of the field does not depend on the sign of the coupling to the Higgs, since it plays a sub-leading role. This allows us to describe the behavior of the field without making any distinction between the two cases. We will first explore the dynamics during the early period of inflation, where the non-minimal coupling to gravity plays the dominant role, and then the subsequent evolution in the radiation era, where the field dynamics is mainly driven by its quartic self-coupling.


\subsection{Inflation}
\label{inflation}

We consider a dark scalar field non-minimally coupled to gravity, with the full action being given by:
\begin{equation}
S=\int \sqrt{-g}\,d^{4}x\left[\frac{1}{2}\,M_{Pl}^{2}\,f\left(\phi\right)R-\frac{1}{2}\,\left(\nabla\phi\right)^{2}-V\left(\phi\right)\right]~,\label{action}
\end{equation}
where $\Phi=\phi/\sqrt{2}$ and $f\left(\phi\right)=1-\xi\,\phi^{2}/M_{Pl}^2$ \cite{foot1}. The equation of motion for the homogeneous field component in a flat FRW Universe is thus:
\begin{equation}
\ddot{\phi}+3H\dot{\phi}+V'\left(\phi\right)+\xi R\phi=0~,\label{eom}
\end{equation}
such that the non-minimal coupling to gravity generates an effective mass for the field. We assume that inflation is driven by some other scalar field, which is consistent since, as we show below, the energy density of $\phi$ is sub-dominant during this phase. We will consider the case where $\xi\gg g,\,\lambda_{\phi}$, such that the dynamics during inflation is mainly driven by the non-minimal coupling to gravity. Since during inflation $R\simeq12\,H_{inf}^{2}$, where the nearly constant Hubble parameter can be written in terms of the tensor-to-scalar ratio $r$:
\begin{equation}
H_{inf}\left(r\right)\simeq2.5\times10^{13}\left(\frac{r}{0.01}\right)^{1/2}\,\mathrm{GeV}~,\label{Hubble infl}
\end{equation}
the effective field mass is:
\begin{equation}
m_{\phi}\simeq\sqrt{12\,\xi}\,H_{inf}~,\label{mass inflation}
\end{equation}
with $m_{\phi}> H_{inf}$ for $\xi>1/12$. 

The mass of the dark scalar has to exceed the Hubble parameter during inflation, since otherwise it develops significant fluctuations on super-horizon scales that may give rise to observable cold dark matter isocurvature modes in the CMB anisotropy spectrum, which are severely constrained by data \cite{Ade:2015lrj}. This implies that the classical field is driven towards the origin during inflation, but its average value can never vanish due to its de-Sitter quantum fluctuations on super-horizon scales. Any massive scalar field exhibits quantum fluctuations that get stretched and amplified by the expansion of the Universe. For $m_{\phi}/H_{inf}>3/2$ ($\xi>3/16$), the amplitude of each super-horizon momentum mode is suppressed by the mass of the dark scalar field, yielding a spectrum \cite{Riotto:2002yw}:
\begin{equation}
\left|\delta\phi_{k}\right|^{2}\simeq\left(\frac{H_{inf}}{2\pi}\right)^{2}\left(\frac{H_{inf}}{m_{\phi}}\right)\frac{2\pi^{2}}{\left(a\,H_{inf}\right)^{3}}~,\label{Fourier modes m greater 3 over 2}
\end{equation}
where $a(t)$ is the scale factor. Integrating over the super-horizon comoving momentum $0<k<aH_{inf}$, at the end of inflation, the homogeneous field variance reads:
\begin{equation}
\left\langle \phi^{2}\right\rangle \simeq\frac{1}{3}\,\left(\frac{H_{inf}}{2\pi}\right)^{2}\frac{1}{\sqrt{12\xi}}~,\label{field variance very massive}
\end{equation}
setting the average amplitude of the field at the onset of the post-inflationary era, $\phi_{inf}$:
\begin{equation}
\phi_{inf}=\sqrt{\left\langle \phi^{2}\right\rangle }\simeq\alpha\,H_{inf}\qquad\alpha\simeq0.05\,\xi^{-1/4}~.\label{initial amplitude after inflation very massive}
\end{equation}
On the other hand, if $\xi<3/16$, $m_{\phi}/H_{inf}<3/2$, the spectrum is given by \cite{Riotto:2002yw}:
\begin{equation}
\left|\delta\phi_{k}\right|\simeq\frac{H_{inf}}{\sqrt{2\,k^{3}}}\,\left(\frac{k}{a\,H_{inf}}\right)^{\frac{3}{2}-\nu_{\phi}}~,\label{fourier modes order Hinf}
\end{equation}
where $\nu_{\phi}=\left(9/4-m_{\phi}^{2}/H_{inf}^{2}\right)^{1/2}$. As above, integrating over all super-horizon modes at the end of inflation, we obtain for the field variance:
\begin{equation}
\left\langle \phi^{2}\right\rangle \simeq\frac{1}{3-2\nu_{\phi}}\left(\frac{H_{inf}}{2\pi}\right)^{2}~,\label{field variance m less than 3 over 2}
\end{equation}
leading to a cold dark matter isocurvature power spectrum of the form:
\begin{equation}
\mathcal{P}_{I}\left(k\right)\simeq\frac{2\pi^{2}}{k^{3}}\,\left(\frac{k}{a\,H_{inf}}\right)^{3-2\nu_{\phi}}\left(3-2\nu_{\phi}\right)~,\label{power spectrum}
\end{equation}
with the corresponding dimensionless power spectrum:
\begin{equation}
\Delta_{I}^{2}\left(k\right)\equiv\frac{k^{3}}{2\,\pi^{2}}\,\mathcal{P}_{I}\left(k\right)=\left(3-2\nu_{\phi}\right)\,\left(\frac{k}{a\,H_{inf}}\right)^{3-2\nu_{\phi}}.\label{dimensionless power spectrum}
\end{equation}
Isocurvature perturbations are bounded by the Planck collaboration in terms of the ratio
\begin{equation}
\beta_{iso}\left(k\right)=\frac{\Delta_{I}^{2}\left(k\right)}{\Delta_{\mathcal{R}}^{2}\left(k\right)+\Delta_{I}^{2}\left(k\right)}~,\label{isocurvature perturbations}
\end{equation}
where $\Delta_{\mathcal{R}}^{2}\simeq2.2\times10^{9}$ is the amplitude of the adiabatic curvature perturbation spectrum generated by the inflaton field. Since the fluctuations in $\phi$ and in the inflaton field are uncorrelated, we may use $\beta_{iso}\left(k_{mid}\right)<0.037$, for $k_{mid}=0.050\,\mathrm{Mpc^{-1}}$, corresponding to the upper
bound on uncorrelated isocurvature perturbations imposed by Planck \cite{Ade:2015lrj}. For 55 e-folds of inflation, this yields $\nu_{\phi}\lesssim1.3$, implying that $m_{\phi}\gtrsim0.75\,H_{inf}$ and setting a lower bound:
\begin{equation}
\xi\gtrsim0.05~,\label{lower bound xi}
\end{equation}
which thus gives for the average homogeneous field amplitude at the end of inflation:
\begin{equation}
\phi_{inf}=\sqrt{\left\langle \phi^{2}\right\rangle }\simeq\alpha\,H_{inf}\qquad\alpha \lesssim 0.25~,\label{initial amplitude case less 3 over 2}
\end{equation}
as found in Ref. \cite{Bertolami:2016ywc}.

Note that the energy-momentum tensor for the dark scalar is given by:
\begin{align}
T_{\mu\nu} & =\left(1-4\xi\right)\nabla_{\mu}\phi\nabla_{\nu}\phi+4\xi\phi\left(g_{\mu\nu}\nabla_{\alpha}\nabla^{\alpha}-\nabla_{\mu}\nabla_{\nu}\right)\phi\nonumber \\
 & \begin{alignedat}{1}+ & g_{\mu\nu}\left[-\left(\frac{1}{2}-4\xi\right)\nabla_{\alpha}\phi\nabla^{\alpha}\phi-V\left(\phi\right)-\xi R\phi^{2}\right]+2\xi\phi^{2}R_{\mu\nu}\end{alignedat}
~,\label{energy-momemtum tensor}
\end{align}
where $R_{\mu\nu}$ is the Ricci tensor. The energy density and pressure of the field are thus, respectively, 
\begin{eqnarray}
\rho_{\phi}&=&\frac{\dot{\phi}^{2}}{2}+V\left(\phi\right)+12\xi H\phi\dot{\phi}+6\xi\phi^{2}H^{2}~,\nonumber\\\label{energy density}
p_{\phi}&  =&\frac{1}{2}\,\left(1-8\xi\right)\dot{\phi}^{2}-V\left(\phi\right)+4\xi\phi V'\left(\phi\right)\
  +4\xi\phi\dot{\phi}H+\xi\phi^{2}\left[\left(8\xi-1\right)R+2\,\frac{\ddot{a}}{a}+4\,H^{2}\right]~.\nonumber\\
\end{eqnarray}
Using $\dot\phi\sim m_\phi \phi$ for an underdamped field, we can easily see that $\rho_\phi \lesssim H_{inf}^4$ for $\xi\lesssim 10$, so that the dark scalar will not affect the inflationary dynamics even for large values of its non-minimal coupling to gravity. Note that this assumes also that the Higgs field does not acquire a large expectation value during inflation, which is natural since de Sitter fluctuations also generate an average Higgs value $\lesssim H_{inf}$ \cite{Enqvist:2014zqa}.

After inflation, the field will oscillate about the minimum of its potential and its effective mass $m_\phi \gg H$, and we show in Appendix \ref{sec:Continuity-equation} that, for $\xi\lesssim 1$, all modifications to the dark field's energy density and pressure due to its non-minimal coupling to gravity become sub-dominant or average out to zero, thus recovering the conventional form for $\xi=0$. In addition, since $R=0$ in a radiation-dominated era and $R\sim \mathcal{O}(H^2)$ in subsequent eras, the non-minimal coupling's contribution to the field's mass also becomes negligible. Thus, we may safely neglect the effects of the non-minimal coupling to gravity in its post-inflationary evolution, hence its only role is to make the field sufficiently heavy during inflation so to prevent the generation of significant isocurvature modes in the CMB anisotropy spectrum.


\subsection{Radiation era}
\label{rad era}

After inflation and the reheating period, which we assume for simplicity to be ``instantaneous", i.e.~sufficiently fast, the Universe undergoes a radiation-dominated era for which $R=0$. Above the electroweak scale, the potential is then dominated by the quartic term,
\begin{equation}
V\left(\phi\right)\simeq\lambda_{\phi}\,\frac{\phi^{4}}{4}~.\label{quartic potential}
\end{equation}
The field is in an overdamped regime until the effective field mass, $m_{\phi}=\sqrt{3\,\lambda_{\phi}}\,\phi$, exceeds the Hubble parameter in this era. At this point, the field begins oscillating about the origin with an amplitude that decays as $a^{-1}\propto T$ and $\rho_{\phi}\sim a^{-4}$, thus behaving like dark radiation.

The temperature at the onset of the post-inflationary field oscillations can be found by equating the effective field mass with the Hubble parameter:
\begin{equation}
T_{rad}=\lambda_{\phi}^{1/4}\,\sqrt{\phi_{inf}\,M_{Pl}}\,\left(\frac{270}{\pi^{2}\,g_{*}}\right)^{1/4}~,\label{T rad}
\end{equation}
where $g_{*}$ is the number of relativistic degrees of freedom. The temperature $T_{rad}$ is thus below the reheating temperature, $T_{rh}\sim\sqrt{M_{Pl}\,H_{inf}}$. Since the field behaves like radiation, its amplitude decreases with $T$:
\begin{equation}
\phi_{rad}  \left(T\right)=\frac{\phi_{inf}}{T_{rad}}\,T
 =\left(\frac{\pi^{2}\,g_{*}}{270}\right)^{1/4}\,\left(\frac{\phi_{inf}}{M_{Pl}}\right)^{1/2}\,\frac{T}{\lambda_{\phi}^{1/4}}~.\label{amplitude field radiation}
\end{equation}
Notice that, above the electroweak scale, thermal effects maintain
the Higgs field localized in the vicinity of the origin, with average
thermal fluctuations $\left\langle h^{2}\right\rangle \ll T^{2}$,
as shown e.g. in Ref. \cite{Bastero-Gil:2016mrl}. We can use this to show that the Higgs-dark scalar
field interactions play a subdominant role before the EWPT. In particular,
in the parametric regime that we are interested in (see Eqs. (\ref{g lambda neg}) and (\ref{g and lambda +})), $g\sim10^{-3}\lambda_{\phi}^{1/4}$, $\xi\lesssim1$ and $r<0.1$:
\begin{equation}
\frac{g^{2}\phi_{rad}^{2}\left\langle h^{2}\right\rangle /4}{\lambda_{\phi}\phi_{rad}^{4}/4}\sim\frac{\left\langle h^{2}\right\rangle }{T^{2}}\ll1,\label{comp quartic quadratic}
\end{equation}
since $\left\langle h^{2}\right\rangle \sim\frac{T^{5}}{M_{Pl}^{3}}$
according to Ref. \cite{Bastero-Gil:2016mrl}.

The dark scalar continues to behave like radiation until the EWPT, at which point the Higgs field acquires its vacuum expectation value, $\mathcal{H}=h/\sqrt{2}= \mathrm{v}/\sqrt{2}$, generating a mass for the dark scalar. The EWPT transition will be completed once the leading thermal contributions to the Higgs potential become Boltzmann-suppressed,  which occurs approximately
at $T_{EW}\sim m_{W}$, where $m_{W}$ is the $W$ boson mass. Comparing the quadratic and quartic terms in the dark scalar potential at $T_{EW}$, we find:
\begin{equation}
\frac{g^2\,\mathrm{v}^{2}\,\phi_{EW}^{2}/4}{\lambda_{\phi}\phi_{EW}^{4}/4}\simeq10^{7}\,\xi^{1/4}\,\left(\frac{r}{0.01}\right)^{-1/2}\,\frac{g^{2}}{\lambda_{\phi}^{1/2}}~,\label{comparison quadratic quartic}
\end{equation}
where we used that $g_{*S}\simeq86.25$ for the number of relativistic degrees of freedom contributing to entropy at $T_{EW}$, and defined $\phi_{EW}\equiv \phi_{rad}(T_{EW})$. Since $r\lesssim 0.1$ \cite{Ade:2015lrj}, in the parametric regime $g\gtrsim10^{-4}\lambda_{\phi}^{1/4}$, $\xi\lesssim1$ , we conclude that the quadratic term is dominant at $T_{EW}$, implying that the field starts behaving as non-relativistic matter already at the EWPT. We will verify explicitly below that this is, in fact, the parametric regime of interest for the field to account for all the present dark matter abundance.

For $T<T_{EW}$ the dynamics of the field is different depending on whether the Higgs-portal coupling is positive or negative, so that we will study these cases separately. We need, however, to ensure that the coherent behaviour of the homogeneous scalar field is preserved until the EWPT, as we explore in detail below.



\subsection{Condensate evaporation}
\label{evaporation}

If its interactions are sufficiently suppressed, the dark scalar field behaves as a long-lived oscillating condensate, i.e, a set of particles with zero (or actually sub-Hubble) momentum that exhibit a collective behavior, and is never in thermal equilibrium with the SM particles. Thereby, we must analyze the constraints on $g$ and $\lambda_{\phi}$
to prevent its thermalization and evaporation into a WIMP-like candidate, the phenomenology of which was studied in Ref. \cite{Enqvist:2014zqa}.
There are two main processes that may lead to condensate evaporation, as we describe in detail below - Higgs annihilation into higher-momentum $\phi$ particles and the perturbative production of $\phi$ particles by the oscillating background field.


\subsubsection{Higgs annihilation into higher-momentum $\phi$ particles}

Higgs bosons in the cosmic plasma may annihilate into $\phi$ pairs, with a rate given by, for $T\gtrsim T_{EW}$:
\begin{equation}
\Gamma_{hh\rightarrow\phi\phi}=n_{h}\left\langle \sigma v\right\rangle ,\label{decay rate}
\end{equation}
where $v\sim c\equiv1$ and $n_{h}$ is the number density of Higgs bosons in the termal bath, 
\begin{equation}
n_{h}=\frac{\zeta\left(3\right)}{\pi^{2}}T^{3}~.\label{Higgs number density thermal bath}
\end{equation}
Before the EWPT, the momentum of the Higgs particles is of the order of the thermal bath temperature, $\left|\mathbf{p}\right|\sim T$, and the cross-section for the process is 
\begin{equation}
\sigma\sim \frac{g^{4}}{64\pi}\,\frac{1}{\left(1+\frac{m_{h}^{2}}{T^{2}}\right)}\,\frac{1}{T^{2}}\,\sqrt{1+\frac{m_{h}^{2}-m_{\phi}^{2}}{T^{2}}}.\label{cross section condensate}
\end{equation}
For $T<T_{EW}$, the Higgs bosons decay into Standard Model degrees of freedom and the production of $\phi$ stops, so that we must require $\Gamma_{hh\rightarrow\phi\phi}\lesssim H$ to prevent the thermalization of the dark scalar condensate for $T>T_{EW}$. Since $\Gamma_{hh\rightarrow\phi\phi}\propto T$ and $H\propto T^{2}$ , the strongest constraint is at $T_{EW}$, which leads to an upper bound on $g$:
\begin{equation}
g\lesssim8\times10^{-4}\,\left(\frac{g_{*}}{100}\right)^{1/8}~,\label{upper bound on g}
\end{equation}
corresponding to an upper bound on the dark scalar's mass $m_\phi\lesssim 100$ MeV.


\subsubsection{Perturbative production of $\phi$ particles by the oscillating background field} \label{perturbative prod}

Another possibility for the condensate's evaporation is the production of $\phi$ particles from the coherent oscillations of the background condensate. Particle production from an oscillating background field in a quartic potential has been studied in detail in Refs. \cite{Kainulainen:2016vzv,Greene:1997fu,Ichikawa:2008ne}. For $T>T_{EW}$, $\phi$ particles are massless and interact with the background field. The coupling between the background field and particle fluctuations, $\delta\phi$, is:
\begin{equation}
\mathcal{L}_{int}=-\frac{3}{2}\,\lambda_{\phi}\,\phi^{2}\,\delta\phi^{2}~.\label{interaction condensate evaporation}
\end{equation}
For a quartic potential, the oscillating condensate evolves as \cite{Ichikawa:2008ne}:
\begin{equation}
\phi\left(t\right)=\frac{\sqrt{\pi}\,\Gamma\left(\frac{3}{4}\right)}{\Gamma\left(\frac{5}{4}\right)}\,\phi_{rad}\,\sum_{n=1}^{\infty}\left(e^{i\left(2n-1\right)\omega t}+e^{-i\left(2n-1\right)\omega t}\right)\,\frac{e^{-\frac{\pi}{2}\left(2n-1\right)}}{1+e^{-\pi\left(2n-1\right)}}~,\label{background field phi}
\end{equation}
with $\omega=\frac{1}{2}\,\sqrt{\frac{\pi}{6}}\,\frac{\Gamma(\frac{3}{4})}{\Gamma(\frac{5}{4})}\,m_{\phi}$ and $m_{\phi}=\sqrt{3\lambda_{\phi}}\,\phi_{rad}$ before the EWPT. The corresponding particle production rate is then given by \cite{Ichikawa:2008ne}:
\begin{equation}
\Gamma_{\phi\rightarrow\delta\phi\delta\phi}\simeq8.86\,\frac{9\,\lambda_{\phi}^{2}}{32\pi\,m_{\phi}^{3}}\,\rho_{\phi}~,\label{decay rate 1}
\end{equation}
The energy density, $\rho_{\phi}$, can be averaged over field oscillations:
\begin{equation}
\rho_{\phi}={1\over2}\langle \dot{\phi}\left(t\right)^{2}\rangle +\frac{\lambda_{\phi}}{4}\,\langle \phi\left(t\right)^{4}\rangle~.\label{energy density condensate evaporation}
\end{equation}
We have computed this numerically, using Eq. (\ref{background field phi}), and obtained:
\begin{equation}
\rho_{\phi}\simeq\frac{1}{4}\,\lambda_{\phi}\,\phi_{rad}^{4}~,\label{energy density condensate}
\end{equation}
implying a particle production rate:
\begin{equation}
\Gamma_{\phi\rightarrow\delta\phi\delta\phi}\simeq4\times10^{-2}\,\lambda_{\phi}^{3/2}\,\phi_{rad}~.\label{decay rate condensate final}
\end{equation}
This is valid for $T>T_{EW}$, whilst after the EWPT the $\phi$ particles acquire mass and the production channel is blocked. Once again, since $\Gamma_{\phi\rightarrow\delta\phi\delta\phi}\propto T$ in a quartic potential, the most stringent constraint is at $T_{EW}$, where $\phi_{rad}=\phi_{EW}$, which places an upper bound on the dark scalar's self-coupling:
\begin{equation}
\lambda_{\phi}<6\times10^{-10}\left(\frac{g_{*}}{100}\right)^{1/5}\left(\frac{r}{0.01}\right)^{-1/5}\xi^{1/10}~.\label{upper bound on lambda}
\end{equation}
If the constraints (\ref{upper bound on g}) and (\ref{upper bound on lambda}) are satisfied, the dark scalar is never in thermal equilibrium with the cosmic plasma, behaving like an oscillating condensate of zero-momentum particles throughout its cosmic history. We will see below that Eq.~(\ref{upper bound on lambda}) yields the most stringent constraint.


\section{Dynamics after the Electroweak symmetry breaking}
\label{after EWSB}

At the EWPT, the field starts behaving differently depending on the sign of the Higgs-portal coupling. If this coupling is negative, the field may acquire a vacuum expectation value and thus become unstable, although sufficiently long-lived to account for dark matter. On the other hand, if the coupling is positive, the U(1) symmetry remains unbroken after the EWPT and the field never decays. In this section, we explore the field dynamics in each case.


\subsection{Negative Higgs-portal coupling}
\label{-}

This case has been explored in Ref. \cite{Cosme:2017cxk}, and here we intend to give a more detailed description of the field dynamics. The relevant interaction potential is:
\begin{equation}
V(\phi,h)=-\,\frac{g^{2}}{4}\,\phi^{2}\,h^{2}+\frac{\lambda_{\phi}}{4}\,\phi^{4}+\frac{\lambda_{h}}{4}\,\left(h^{2}-\tilde{v}^{2}\right)^{2}~.\label{Lagrangian neg}
\end{equation}
As soon as the temperature drops below the electroweak scale, both the Higgs field and the dark scalar acquire non-vanishing vacuum expectation values for $g^4<4\lambda_\phi \lambda_h$, which we take to be the relevant parametric regime:
\begin{equation}
h_{0}=\left(1-\frac{g^{4}}{4\,\lambda_{\phi}\,\lambda_{h}}\right)^{-1/2}\tilde{v}\equiv\mathrm{v},\qquad\phi_{0}=\frac{g\,\mathrm{v}}{\sqrt{2\lambda_{\phi}}}~,\label{vevs}
\end{equation}
where $\mathrm{v}=246\,\mathrm{GeV}$. This induces a mass-mixing between the ``flavour" basis $\phi$ and $h$ fields, described by the squared mass matrix
\begin{equation}
M^{2}=\mathrm{v}^{2}\begin{pmatrix}g^{2} & -\frac{g^{3}}{\sqrt{2\lambda_{\phi}}}\\
-\frac{g^{3}}{\sqrt{2\lambda_{\phi}}} & 2\lambda_{h}
\end{pmatrix}.\label{mass mixing matrix}
\end{equation}
For small mixing, the mass eigenvalues are approximately given by:
\begin{equation}
m_{\tilde{\phi}}^{2}  \simeq g^{2}\mathrm{v}^{2}\left(1-\frac{g^{4}}{4\,\lambda_{h}\,\lambda_{\phi}}\right)~,\qquad m_{\tilde{h}}^{2}  \simeq2\lambda_{h}\,\mathrm{v}^{2}\left(1+\frac{g^{6}}{8\,\lambda_{h}^{2}\,\lambda_{\phi}}\right)~.\label{m phi squared correc}
\end{equation}
The corresponding eigenvectors are:
\begin{equation}
\tilde{\phi}=\begin{pmatrix}1\\
\epsilon_{-}
\end{pmatrix},\qquad\tilde{h}=\begin{pmatrix}-\epsilon_{-}\\
1
\end{pmatrix}~,\label{phi h flavour basis}
\end{equation}
where the mixing parameter, $\epsilon_{-}\ll 1$, corresponds to
\begin{align}
\epsilon_{-} & =\frac{g^{3}}{2\,\sqrt{2\,\lambda_{\phi}}\,\lambda_{h}}=\frac{g^{2}\,\phi_{0}\,\mathrm{v}}{m_{h}^{2}}~.\label{epsilon}
\end{align}
The mass eigenstates can then be written in terms of the flavour-basis fields as:
\begin{equation}
\tilde{\phi} =\phi-\epsilon_{-}\,h ~,\qquad \tilde{h}  =h+\epsilon_{-}\,\phi~.
\label{mixing h and phi}
\end{equation}
This mixing is extremely relevant for the direct and indirect detection of the dark scalar as we discuss in section \ref{laboratory}. In this scenario, differentiating
the Lagrangian twice with respect to $\phi$, yields the mass of the field:
\begin{equation}
m_{\phi}=g\,\mathrm{v}~.\label{mass neg gv}
\end{equation}
As we have seen in section \ref{rad era}, the field starts behaving as cold dark matter at $T_{EW}\sim m_{W}$. The amplitude of the oscillations at this stage is given by Eq. (\ref{amplitude field radiation}), which can be rewritten as:
\begin{align}
\phi_{EW} & \simeq\left(\frac{4\,\pi^{2}\,g_{*}}{270}\right)^{1/4}\,\left(\frac{\phi_{inf}}{M_{Pl}}\right)^{1/2}\,\frac{T_{EW}}{\mathrm{v}}\,\frac{\lambda_{\phi}^{1/4}}{g}\,\phi_{0}\nonumber \\
 & \simeq10^{-4}\,g_{*}^{1/4}\,\xi^{-1/8}\!\left(\frac{T_{EW}}{m_{W}}\right)\left(\frac{r}{0.01}\right)^{1/4}\,\frac{\lambda_{\phi}^{1/4}}{g}\,\phi_{0}.\label{phi EW negative}
\end{align}
From Eq. (\ref{phi EW negative}), in the parametric regime $g\gtrsim10^{-4}\,\lambda_{\phi}^{1/4}$, $\phi_{EW}\lesssim\phi_{0}$, for $\xi\gtrsim1$. This implies that the potential minimum will move smoothly from the origin to $\phi_{0}$, where the field starts to oscillate with an amplitude $\phi_{DM}=x_{DM}\,\phi_{0}$, with $x_{DM}\lesssim1$.
In fact, a rigorous study of the dynamics of the field for $T_{EW}<T<T_{CO}$, where $T_{CO}$ corresponds to the electroweak crossover temperature,
would necessarily involve numerical simulations that are beyond the scope of this paper. We may nevertheless estimate the uncertainty associated to the field's amplitude. Since $T_{EW}\lesssim T_{CO}$ by an $\mathcal{O}\left(1\right)$ factor, and given that $\phi\sim T$ while behaving as radiation and $\phi\sim T^{3/2}$ while behaving as non-relativistic
matter, the field's amplitude might decrease by at most an $\mathcal{O}\left(1\right)$ factor as well. Therefore, we expect the field's amplitude to be smaller than
$\phi_{0}$ at $T_{EW}$ by $x_{DM}\lesssim 1$. Notice that $x_{DM}$ is thus not an additional parameter of the model, but only a theoretical uncertainty in our
analysis that does not affect the order of magnitude of the dark matter abundance and lifetime, as we shall see below.

As soon as the field reaches $\phi_{0}$, it stops behaving like dark radiation and starts to oscillate as cold dark matter. The field's amplitude of oscillations about $\phi_0$ evolves according to
\begin{equation}
\phi\left(T\right)=\phi_{DM}\left(\frac{T}{T_{EW}}\right)^{3/2}~.\label{amplitude CDM neg}
\end{equation}
At this point, the number of particles in a comoving volume becomes constant:
\begin{equation}
\frac{n_{\phi}}{s}=\frac{45}{4\pi^{2}g_{*S}}\frac{m_{\phi}\,\phi_{DM}^{2}}{T_{EW}^{3}}~,\label{n over s neg}
\end{equation}
where $s=\frac{2\pi^{2}}{45}\,g_{*S}\,T^{3}$ is the entropy density of radiation, $n_{\phi}\equiv\frac{\rho_{\phi}}{m_{\phi}}$, is the dark matter number density and $g_{*S}\simeq86.25$ at $T_{EW}$. Using this, we can compute the present dark matter abundance, $\Omega_{\phi,0}\simeq0.26$, obtaining the following relation between the latter and $m_{\phi}$:
\begin{equation}
m_{\phi}=\left(6\,\Omega_{\phi,0}\right)^{1/2}\left(\frac{g_{*S}}{g_{*S0}}\right)^{1/2}\left(\frac{T_{EW}}{T_{0}}\right)^{3/2}\frac{H_{0}M_{Pl}}{\phi_{DM}},\label{mass neg}
\end{equation}
where $g_{*S0}$, $T_{0}$ and $H_{0}$ are the present values of the number of relativistic degrees of freedom, CMB temperature and
Hubble parameter, respectively. We thus obtain the following relation between the couplings $g$ and $\lambda_{\phi}$:
\begin{equation}
g\simeq2\times10^{-3}\,\left(\frac{x_{DM}}{0.5}\right)^{-1/2}\,\lambda_{\phi}^{1/4}~.\label{g lambda neg}
\end{equation}
This expression satisfies the constraint $g^4<4\lambda_\phi\lambda_h$ and $\phi_{EW}\lesssim\phi_{0}$. Given this relation between couplings, Eq. (\ref{upper bound on lambda}) yields, as anticipated, the strongest constraint, limiting the viable dark matter mass to be  $\lesssim1\,\mathrm{MeV}$.

It should be emphasized that the properties of our model depend strongly whether the U(1) symmetry is global or local. In particular, if the symmetry is global, it leads to tight constraints on the dark scalar mass. We study the cosmological implications of the spontaneous symmetry breaking (both local and global) in section \ref{Cosmo implications}.


\subsection{Positive Higgs-portal coupling}
\label{+}

As we have seen before, for $T>T_{EW}$, the field is oscillating as radiation. If the coupling between the Higgs and the SDFM has a positive sign, as soon as the EWPT occurs, the only field that undergoes spontaneous symmetry breaking is the Higgs field. The corresponding interaction Lagrangian is:
\begin{equation}
\mathcal{L}_{int}=+\,\frac{g^{2}}{4}\,\phi^{2}\,\mathrm{v}^{2}+\frac{\lambda_{\phi}}{4}\,\phi^{4}+\frac{\lambda_{h}}{4}\,\left(h^{2}-\mathrm{v}^{2}\right)^{2}~.\label{Lagrangian pos}
\end{equation}
The interactions with the Higgs are responsible for providing the dark scalar mass, which is given in this case by
\begin{equation}
m_{\phi}=\frac{1}{\sqrt{2}}\,g\,\mathrm{v}~,\label{mass pos}
\end{equation}
differing from the mass of the negative coupling case, Eq. (\ref{mass neg gv}), by a factor of $\sqrt{2}$ . Although there is no spontaneous U(1) symmetry breaking, there is in practice an effective mass-mixing between the Higgs and dark scalar fields, given that the latter's value is oscillating about the origin but does not vanish exactly, yielding an effective mixing parameter:
\begin{equation}
\epsilon_{+}=\frac{g^{2}\,\phi\,\mathrm{v}}{m_{h}^{2}}~,\label{epsilon NB 1}
\end{equation}
where $\phi$ denotes the time-dependent field amplitude. Since the dark matter energy density depends on the amplitude of the field,
\begin{equation}
\rho_{\phi}=\frac{1}{2}\,m_{\phi}^{2}\,\phi^{2}~,\label{present DM abundance}
\end{equation}
Eq. (\ref{epsilon NB 1}) can be rewritten as
\begin{equation}
\epsilon_{+}=\frac{2\,\sqrt{2\,\rho_{\phi}}}{\mathrm{v}\,m_{h}^{2}}\,m_{\phi}~.\label{epsilon nb2}
\end{equation}
At $T_{EW}$, the field starts to behave as cold dark matter, and its amplitude varies as:
\begin{equation}
\phi\left(T\right)=\frac{\phi_{inf}}{T_{rad}}\,\left(\frac{T^{3}}{T_{EW}}\right)^{1/2}~,\label{amplitude matt}
\end{equation}
so that the number of particles per comoving volume reads:
\begin{equation}
\frac{n_{\phi}}{s}=\frac{45}{4\pi^{2}}\,\frac{1}{g_{*S}}\,m_{\phi}\,\frac{\phi_{inf}^{2}}{T_{rad}^{2}}\,\frac{1}{T_{EW}}~.\label{comoving volume}
\end{equation}
Analogously to the negative coupling case, we may compute the field's mass, taking into account the present dark matter abundance, $\Omega_{\phi,0}$:
\begin{equation}
m_{\phi}=\sqrt{6\,\Omega_{\phi,0}}\,\frac{H_{0}\,M_{Pl}}{\phi_{inf}}\,\left(\frac{g_{*S}}{g_{*S0}}\right)^{1/2}\,\frac{T_{rad}\,T_{EW}^{1/2}}{T_{0}^{3/2}}~,\label{mass +}
\end{equation}
leading to the following relation between $g$ and $\lambda_{\phi}$:
\begin{equation}
g\simeq9\times10^{-3}\,\xi^{1/8}\,\left(\frac{r}{0.01}\right)^{1/4}\,\lambda_{\phi}^{1/4}~.\label{g and lambda +}
\end{equation}
Once again, Eq. (\ref{upper bound on lambda}) yields the strongest constraint and, as in the case of the negative coupling, the dark scalar's mass must be smaller than 1 MeV to account for the present dark matter abundance.

It must be highlighted that, in this scenario, once the dark matter abundance is fixed, the value of $g$ depends on all other parameters of the model, $\lambda_{\phi}$, $r$ and $\xi$, contrary to the negative coupling case, where the Higgs-portal coupling is only related to $\lambda_{\phi}$  as given in Eq. (\ref{g lambda neg}). In other words, in the positive coupling case the field's mass depends on the initial conditions set by inflationary dynamics, whilst in the case with spontaneous symmetry breaking the final dark matter abundance is effectively independent of the initial conditions, and all dynamics depends essentially on the vacuum expectation value, $\phi_{0}$, which sets the amplitude of field oscillations.


%
\section{Phenomenology}
\label{Pheno}

Having discussed the dynamics of the dark scalar throughout the cosmic history and determined the parametric ranges for which it may account for the present dark matter abundance, in this section we discuss different physical phenomena that may be used to probe the model. This includes searching for direct signatures in the laboratory, as well as for indirect signals in astrophysical observations.


\subsection{Astrophysical Signatures}

Given the smallness of the Higgs portal coupling required for preventing the scalar condensate's evaporation before the EWPT, it is likely that any signatures of such a dark matter candidate require large fluxes of $\phi$ particles or the particles it interacts with, which may be difficult to produce in the laboratory. Astrophysical signatures may, however, be enhanced by the large dark matter density within our galaxy and other astrophysical systems, and thus we shall explore in this section the possibility of indirectly detecting the dark scalar through its annihilation or decay.


\subsubsection{Dark matter annihilation}

Over the past decades, the emission of a $511$ keV $\gamma$-ray line has been observed, by several experiments, in a region around the galactic center (see e.g.~Ref. \cite{Prantzos:2010wi} for a review). The $511$ keV line is characteristic of electron-positron annihilation, as has been shown by the INTEGRAL/SPI observations \cite{Jean:2005af,Knodlseder:2005yq,Churazov:2004as} and has a flux of $\Phi_{GC}=9.9\times10^{-4}\,\mathrm{cm^{-2}}\,s^{-1}$ \cite{Jean:2003ci}. This photon excess in the galactic bulge is intriguing because, although it is common to find positron sources in the galaxy (namely, from core collapse of supernovae and low-mass X-ray binaries), most of these astrophysical objects are localized in the galactic disk rather than in the galactic center \cite{Prantzos:2010wi}.

An alternative possibility to explain this line is considering dark matter annihilation into an $e^{-}e^{+}$ pair. Although light WIMPs are practically excluded as a possible explanation \cite{Wilkinson:2016gsy}, other plausible dark matter candidates could predict this photon excess. In this context, we study the annihilation of $\phi$ particles into
$e^{-}e^{+}$ and into photons, presenting the predictions for the photon flux associated with our dark matter candidate in the galactic center.

In general, the flux of photons from an angular region $\Delta\Omega$ from dark matter annihilation is given by:
\begin{equation}
\Phi_{ann}=\frac{1}{2m_{\phi}^{2}}N_{\gamma}\frac{\left\langle \sigma v\right\rangle }{4\pi}\int_{l.o.s}dl\left(s,\Psi\right)\,\rho^{2}\left(r\left(s,\Psi\right)\right)\int_{\Delta\Omega}d\Omega~,\label{flux ann}
\end{equation}
where $\rho$ is the dark matter density for a given profile, $r$ is the radial distance from the galactic center, $\Psi$ is the angle between the direction of observation in the sky and the galactic center, the integral is evaluated along the line-of-sight (l.o.s) and $N_{\gamma}$ is the number of photons produced by the annihilation \cite{Bergstrom:1997fj}. We
can split Eq.~(\ref{flux ann}) into two parts: one that only depends on the particle physics,
\begin{equation}
\frac{1}{2m_{\phi}^{2}}N_{\gamma}\frac{\left\langle \sigma v\right\rangle }{4\pi}~,\label{Particle Physics dependence}
\end{equation}
and another one depending only on the astrophysical properties:
\begin{equation}
J\left(\Omega\right)=\int_{l.o.s}dl\left(s,\Psi\right)\,\rho^{2}\left(r\left(s,\Psi\right)\right)\int_{\Delta\Omega}d\Omega~,\label{J factor}
\end{equation}
and analyze each one separately. 

Let us first focus on the particle physics component. Consider the case in which the $\phi$ particles annihilate via virtual Higgs exchange, $H^{*}$, and the latter decays into Standard Model particles. Since $m_{\phi}\lesssim1$ MeV (by the constraint imposed on $\lambda_{\phi}$, Eq. (\ref{upper bound on lambda})), the possible final decay products are only electron-positron pairs and photons for non-relativistic dark scalars. Following Eq. (\ref{Particle Physics dependence}), it is necessary to compute the cross section of the process. In the center-of-mass frame, using $E_{\phi}\simeq m_{\phi}$, the 4-momentum of $H^{*}$ is $p_{h}=\left(2\,m_{\phi},\mathbf{0}\right)$, implying that the invariant mass of the virtual Higgs is $m_{H^{*}}=2\,m_{\phi}$. By energy conservation, if the final state is a fermion-antifermion pair, the momentum of each particle reads $\left|\mathbf{p}_f\right|=m_{\phi}\sqrt{1-m_f^2/m_\phi^2}$, where $m_{f}$ is the fermion's mass. Using this, the cross section for dark matter annihilation into a virtual Higgs and its subsequent decay into fermions reads:
\begin{equation}
\left\langle \text{\ensuremath{\sigma}}v_{rel}\right\rangle _{fermions}\simeq\frac{N_{c}}{8\pi}\,\frac{D}{\mathrm{v^{2}}}\,\frac{m_{\phi}^{4}}{\left(4m_{\phi}^{2}-m_{h}^{2}\right)^{2}+m_{h}^{2}\Gamma_{h}^{2}}\,\left(\frac{m_{f}}{\mathrm{v}}\right)^{2}\,\left(1-\frac{m_{f}^{2}}{m_{\phi}^{2}}\right)^{3/2}~,\label{cross section fermions}
\end{equation}
where $N_{c}$ is the number of colors ($N_{c}=1$ for leptons and $N_{c}=3$ for quarks), $\Gamma_{h}=4.07\times10^{-3}\,\mathrm{GeV}$ \cite{Cuoco:2016jqt} is the total Higgs decay rate, and $D=1$ for the case where the U(1) symmetry is spontaneously broken and $D=4$ otherwise.

In the case where we have a pair of photons in the final state, the momentum of each photon is $\left|\mathbf{p}_{\gamma}\right|=m_{\phi}$. The cross-section for this process is:
\begin{equation}
\left\langle \text{\ensuremath{\sigma}}v_{rel}\right\rangle _{\gamma}\simeq\frac{D}{32\sqrt{2}\,\pi^{3}}\,\frac{m_{\phi}^{4}}{\mathrm{v}^{2}}\,\frac{G_{F}\,\alpha_{QED}^{2}\,m_{\phi}^{2}}{\left(4m_{\phi}^{2}-m_{h}^{2}\right)^{2}+m_{h}^{2}\Gamma_{h}^{2}}\,F^{2}~,\label{cross section photons}
\end{equation}
where $G_{F}=1.17\times10^{-5}\,\mathrm{GeV}^{-2}$ is Fermi's constant, $\alpha_{QED}\simeq1/137$ is the fine structure constant
and 
\begin{equation}
F=\left|\sum_{f}\,N_{c}\,Q_{f}^{2}\,A_{1/2}^{H}\left(\tau_{f}\right)+A_{1}^{H}\left(\tau_{\mathbf{w}}\right)\right|\simeq\frac{11}{3}\label{F factor}
\end{equation}
is a factor that takes into account the loop contributions of all charged fermions and the $W$ boson to $H^*\rightarrow \gamma\gamma$, with $\tau_{i}=4m_{i}^{2}/m_{\phi}^{2}\gg1$ for all particle species involved, for $m_{\phi}\ll\mathrm{MeV}$ \cite{Djouadi:2005gi}. We point out that for the decay of a virtual Higgs all charged fermions give essentially the same loop contribution, whilst for an on-shell Higgs boson only the top quark contributes significantly.

Concerning the astrophysical part, one needs to consider the profile that best describes the dark matter distribution in a particular region of the Universe to compute the so-called $J$-factor in Eq. (\ref{J factor}). In the case of the galactic center, the most appropriate models are the cuspy profile ones. Following Ref. \cite{Cuoco:2016jqt}, using a generalized Navarro-Frenk-White profile (NFW) to describe the dark matter distribution in the galactic center,
\begin{equation}
\rho\left(r\right)=\rho_{s}\,\left(\frac{r}{r_{s}}\right)^{-\gamma}\,\left(1+\frac{r}{r_{s}}\right)^{-3+\gamma}~,\label{generalized NFW}
\end{equation}
where $r$ is the spherical distance from the galactic center, $\gamma=1.2$, $\rho_{s}=0.74\,\mathrm{GeV/cm^{3}}$ and $r_{s}=19.5$ kpc; hence, the corresponding $J$-factor is:
\begin{equation}
J\simeq1.79\times10^{23}\,\mathrm{GeV^{2}\,cm^{-5}}.\label{Jfactor galactic center}
\end{equation}
Using the last expressions, the predicted flux of photons originated by the annihilation of $\phi$ particles into a virtual Higgs and the decay of the latter into an electron-positron pair, for the galactic center, is given by:
\begin{equation}
\Phi_{\phi\phi\rightarrow e^{-}e^{+}}\simeq\frac{D}{32\,\pi^{2}}\,\frac{J}{\left(4m_{\phi}^{2}-m_{h}^{2}\right)^{2}+m_{h}^{2}\Gamma_{h}^{2}}\,\left(\frac{m_{\phi}}{\mathrm{v}}\right)^{2}\,\left(\frac{m_{e}}{\mathrm{v}}\right)^{2}\,\left(1-\frac{m_{e}^{2}}{m_{\phi}^{2}}\right)^{3/2}~.\label{flux phi electrons}
\end{equation}
Since $m_{\phi}\lesssim1$ MeV and $m_{\phi}\gtrsim0.511$ MeV so that it can annihilate into an electron-positron pair, we can simplify this to give:
\begin{equation}
\Phi_{\phi\phi\rightarrow e^{-}e^{+}}\simeq1\times10^{-27}\,D\,\left(\frac{m_{\phi}}{\mathrm{MeV}}\right)^{2}\,\mathrm{cm^{-2}}\,s^{-1}~,\label{flux phi elect final}
\end{equation}
which is clearly insufficient to explain the galactic center excess but shows how the electron-positron flux varies with the mass of the dark scalar and hence the Higgs-portal coupling. 

The flux coming from the annihilation of $\phi$ into a virtual Higgs and its decay into photons reads:
\begin{equation}
\Phi_{\phi\phi\rightarrow\gamma\gamma}\simeq10^{-31}\,D\,\left(\frac{m_{\phi}}{\mathrm{MeV}}\right)^{4}\,\mathrm{cm^{-2}}\,s^{-1}~.\label{flux photons}
\end{equation}
The fluxes we have found for the annihilation of $\phi$ particles into a virtual Higgs and its decay into an electron-positron pair or photons are thus extremely suppressed, which makes its indirect detection through annihilation a very difficult challenge.


\subsubsection{Dark matter decay}

Our dark matter candidate exhibits a small mixing $\epsilon_\pm$ with the Higgs boson, as we have seen in section \ref{after EWSB} and may, therefore, decay into the same channels as the Higgs boson, provided that they are kinematically accessible, and the decay width is suppressed by a factor $\epsilon_\pm^{2}$ relative to the corresponding Higgs partial width, $\Gamma_{\phi\rightarrow pp}=\epsilon_\pm^{2}\,\Gamma_{H^{*}\rightarrow XX}$, where $X$ corresponds to a generic particle. The partial decay width of a virtual Higgs boson into photons is \cite{Djouadi:2005gi}:
\begin{equation}
\Gamma_{H^{*}\rightarrow\gamma\gamma}=\frac{G_{F}\,\alpha_{QED}^{2}\,m_{\phi}^{3}}{128\sqrt{2}\,\pi^{3}}F^{2}~,\label{higgs photons}
\end{equation}
where $F$ is defined in Eq. (\ref{F factor}). This yields for the dark scalar's lifetime: 
\begin{equation}
\tau_{\phi}^{\left(-\right)}\simeq7\times10^{27}\left(\frac{7\,\mathrm{keV}}{m_{\phi}}\right)^{5}\,\left(\frac{x_{DM}}{0.5}\right)^{2}\,\mathrm{sec}~,\label{lifetime B}
\end{equation}
in the case with spontaneous symmetry breaking and
\begin{equation}
\tau_{\phi}^{\left(+\right)}\simeq4\times10^{71}\left(\frac{7\,\mathrm{keV}}{m_{\phi}}\right)^{5}\,\mathrm{sec}~,\label{lifetime NB}
\end{equation}
for the positive coupling. As expected, the case with no spontaneous symmetry breaking yields an effectively stable dark matter candidate, with the probability of decay into photons being tremendously suppressed. In the negative coupling case, the dark scalar's lifetime is larger than the age of the Universe, but there is a realistic possibility of producing an observable monochromatic line in the galactic spectrum.

In fact, the XMM-Newton X-ray observatory has recently detected a line at $3.5$ keV in the galactic center, Andromeda and Perseus cluster \cite{Bulbul:2014sua,Boyarsky:2014jta,Boyarsky:2014ska,Cappelluti:2017ywp}. The origin of the line is not well-established, although dark matter decay and/or annihilation are valid possibilities \cite{Higaki:2014zua,Jaeckel:2014qea,Dudas:2014ixa,Queiroz:2014yna,Cappelluti:2017ywp, Heeck:2017xbu}. There are other astrophysical processes that might explain the $3.5$ keV line \cite{Jeltema:2014qfa}. However, some independent studies suggest that those processes cannot provide a satisfactory explanation \cite{Boyarsky:2014paa,Bulbul:2014ala,Iakubovskyi:2015kwa}. In addition, some groups state that such a photon excess is not present in dwarf galaxies, such as Draco \cite{Jeltema:2015mee}, while others assert that the line is only too faint to be detected with the technology that we have at our current disposal, not ruling out a possible dark matter decay interpretation \cite{Ruchayskiy:2015onc}.

According to Refs. \cite{Boyarsky:2014ska,Ruchayskiy:2015onc}, the line present in the galactic center, Andromeda and Perseus can be explained by the decay of a dark matter particle
with a mass of $\sim7$ keV with a lifetime within the interval $\tau_{\phi}\sim\left(6-9\right)\times10^{27}$ sec, which also explains the absence of such a line in the blank-sky data
set. Concerning the positive coupling scenario, the dark scalar's lifetime (Eq.~(\ref{lifetime NB})) is not compatible with the required range. However, the scenario where $\phi$ undergoes spontaneous symmetry breaking yields a lifetime (Eq. (\ref{lifetime B})) that matches the above-mentioned range, up to an uncertainty on the value of the field amplitude after the EWPT parametrized by $x_{DM}\lesssim1$, for $m_{\phi}=7\,\mathrm{keV}$. Furthermore, for this mass, we have $g\simeq3\times10^{-8}$ and, from Eq. (\ref{g lambda neg}), $\lambda_{\phi}\simeq4\times10^{-20}$, satisfying the constraints in Eqs. (\ref{upper bound on g}) and (\ref{upper bound on lambda}), as we had already presented in Ref. \cite{Cosme:2017cxk}.

We should emphasize the uniqueness of the result for the spontaneous symmetry breaking scenario. In this case, our model predicts that the decay of $\phi$ into photons produces a 3.5 keV line compatible with the observational data, with effectively only a single free parameter, either $g$ or $\lambda_{\phi}$. Recall that the initial
conditions of the field depend on the parameters $r$ and $\xi$, the first one determining the initial oscillation amplitude in the radiation era, whilst the latter is responsible for suppressing cold dark matter isocurvature perturbations. However, after the EWPT, the field oscillates around $\phi_{0}$, which only depends on $g$ and $\lambda_{\phi}$. Its amplitude is of the order of $\phi_{0}$, meaning that $r$ and $\xi$ do not play a significant role below $T_{EW}$. Therefore, we have three observables that rely on just two parameters ($g$ and $\lambda_{\phi}$) -  the present dark matter abundance, the dark scalar's mass and its lifetime. However, assuming that the dark scalar field accounts for all the dark matter in the Universe imposes a relation between $g$ and $\lambda_{\phi}$ (Eq.~(\ref{g lambda neg})), implying that $m_{\phi}$ and $\tau_{\phi}$ depend exclusively on the Higgs-portal coupling. Hence, the prediction for the magnitude of the 3.5 keV line in different astrophysical objects is quite remarkable and, as far as we are aware, it has not been achieved by other scenarios, where the dark matter's mass and lifetime can be tuned by different free parameters.


\subsection{Laboratory Signatures}
\label{laboratory}

Since $m_{\phi}\lesssim\mathrm{MeV}$, the coupling between the dark scalar and the Higgs boson, $g$, is extremely small, which hampers the detection of dark matter candidates of this kind. However, the small $h-\phi$ mass mixing may nevertheless lead to interesting experimental signals that one could hope to probe in the laboratory with improvements in current technology. In this section, we explore examples of such signatures, namely invisible Higgs decays, photon-dark scalar oscillations in an external electromagnetic field and dark matter-induced oscillations of fundamental constants, such as the electron mass and $\alpha_{QED}$.


\subsubsection{Invisible Higgs decays into dark scalars}
The simplest and cleanest way to test the Higgs-portal scalar field dark matter scenario in collider experiments is to look for invisible Higgs decays into dark scalar pairs, with a decay width:
\begin{equation}
\Gamma_{h\rightarrow\phi\phi}=\frac{1}{8\pi}\,\frac{g^{4}\mathrm{v^{2}}}{4\,m_{h}}\,\sqrt{1-\frac{4m_{\phi}^{2}}{m_{h}^{2}}}~,\label{decay H inv}
\end{equation}
where $m_{h}$ is the Higgs mass. The current experimental limit on the branching ratio for Higgs decay into invisible particles is \cite{Aad:2015pla}:
\begin{equation}
Br\left(\Gamma_{h\rightarrow inv}\right)=\frac{\Gamma_{h\rightarrow inv}}{\Gamma_{h}+\Gamma_{h\rightarrow inv}}\lesssim0.23~.\label{branching ratio}
\end{equation}
Assuming that $\Gamma_{h\rightarrow inv}=\Gamma_{h\rightarrow\phi\phi}$ and using $\Gamma_{h}=4.07\times10^{-3}$ GeV \cite{Cuoco:2016jqt},
this yields the following bound on the Higgs-portal coupling $g$:
\begin{equation}
g\lesssim0.13~,\label{constraint inv}
\end{equation}
which is much less restrictive than the cosmological bound in Eq. (\ref{upper bound on g}). Conversely, for $m_{\phi}<\mathrm{MeV}$ to ensure the survival of the dark scalar condensate up to the present day, we obtain the following upper bound on the branching ratio:
\begin{equation}
Br\left(\Gamma_{h\rightarrow inv}\right)<10^{-19}~.\label{BR MeV}
\end{equation}
This is, unfortunately, too small to be measured with current technology, but may serve as motivation for extremely precise measurements of the Higgs boson's width in future collider experiments, given any other experimental or observational hints for light Higgs-portal scalar field dark matter, such as, for instance, the 3.5 keV line discussed above.


\subsubsection{Light Shining Through Walls}

The dark scalar exhibits a small coupling to photons through its mass mixing with the Higgs boson, which couples indirectly to the electromagnetic field via $W$-boson and charged fermion loops \cite{Djouadi:2005gi}. This will allow us to explore experiments that make use of the coupling to the electromagnetic field to probe dark matter candidates, as the case of ``light shining through a wall'' experiments (LSTW). These experiments are primarily designed to detect WISPs - Weakly Interacting Sub-eV Particles, such as axions and axion-like particles (ALPs), taking advantage of their small coupling to photons. Therefore, we intend to apply the same detection principles to our dark scalar field, given its similarities with ALPs, in terms of small masses and couplings. 

In LSTW experiments, photons are shone into an opaque absorber, and some of them may be converted into WISPs, which traverse the absorber wall. Behind the wall, some WISPs reconvert back into photons that may be detected. This can be achieved using an external magnetic (or electric) field, which works as a mixing agent, making the photon-WISP oscillations possible and allowing for the conversion of photons into WISPs and vice-versa. Using the paradigmatic case of a pseudo-scalar axion/ALP, the interaction term between the latter and photons is of the form:
\begin{equation}
\mathcal{L}_{a\gamma\gamma}=\frac{g_{a\gamma\gamma}}{4}\,a\,F_{\mu\nu}\tilde{F}^{\mu\nu}=-g_{a\gamma\gamma}\,a\,\mathbf{E}.\mathbf{B}~,\label{axion photon photon int term}
\end{equation}
where $F_{\mu\nu}$ is the electromagnetic field tensor, $\tilde{F}^{\mu\nu}$ its dual and $a$ the axion/ALP field. For the QCD axion, the coupling $g_{a\gamma\gamma}$ reads
\begin{align}
g_{a\gamma\gamma} & =\frac{\alpha_{QED}\,K}{2\pi\,F_{a}}
  \simeq2\times10^{-15}K\left(\frac{m_{a}}{10^{-5}\,\mathrm{eV}}\right)\,\mathrm{GeV^{-1}}~,\label{coupling axion photon photon}
\end{align}
with $F_{a}\simeq6\times10^{11}\,\left(\frac{m_{a}}{10^{-5}\,\mathrm{eV}}\right)^{-1}$ GeV denoting the axion decay constant, $m_{a}$ the axion mass and $K\sim\mathcal{O}\left(1-10\right)$ is a model-dependent factor \cite{Adler:1969gk,Cheng:1995fd,Rosa:2017ury}. For scalar WISPs the interaction with the electromagnetic field is given by \cite{Redondo:2010dp,Ehret:2010mh}:
\begin{equation}
\mathcal{L}_{a\gamma\gamma}=\frac{g_{a\gamma\gamma}}{4}\,a\,F_{\mu\nu}F^{\mu\nu}=g_{a\gamma\gamma}\,a\,\left(\mathbf{B}^{2}-\mathbf{E}^{2}\right)~.\label{int 2 axion photon photon}
\end{equation}
In both cases, the presence of an external magnetic field, $\mathbf{B}_{ext}$, gives rise to a mass mixing term between $a$ and photons, thus inducing photon-ALPs oscillations. In the relativistic limit, the WISP wavenumber is $k_{a}\thickapprox\omega-\frac{m_{a}^{2}}{2\omega}$, where $\omega$ is the photon angular frequency \cite{Redondo:2010dp}.
Using this, considering the conversion of a photon into a WISP in a constant magnetic field of length $L$, in a symmetric LSTW setup, the conversion probability of a photon into a WISP, $P_{\gamma\rightarrow a}$, is the same as for the inverse process, $P_{a\rightarrow\gamma}$, being given by \cite{Redondo:2010dp,Ehret:2010ki}:
\begin{equation}
P_{\gamma\rightarrow a}=4\,\frac{g_{a\gamma\gamma}^{2}\,B_{ext}^{2}\,\omega^{2}}{m_{a}^{4}}\,\sin^{2}\left(\frac{m_{a}^{2}\,L}{4\,\omega}\right)~.\label{prob gamma a}
\end{equation}
Therefore, the total conversion probability along the path is simply the square of Eq. (\ref{prob gamma a}) \cite{Ehret:2010ki}:
\begin{equation}
P_{\gamma\rightarrow a\rightarrow\gamma}=16\,\frac{g_{a\gamma\gamma}^{4}\,B_{ext}^{4}\,\omega^{4}}{m_{a}^{8}}\,\sin^{4}\left(\frac{m_{a}^{2}\,L}{4\,\omega}\right)~.\label{prob total}
\end{equation}
In the specific case of the QCD axion, for $L\ll 4\omega/m_a^2$, the LSTW probability is approximately given by:
\begin{equation}
P_{\gamma\rightarrow a\rightarrow\gamma}=6\times10^{-59}\,\left(\frac{B_{ext}}{1\ \mathrm{T}}\right)^{4}\left({m_a\over 10^{-5}\ \mathrm{eV}}\right)^{4}\left(\frac{L}{1\ \mathrm{m}}\right)^4~,\label{total prob axion}
\end{equation}
and, using the parameters of the LSTW experiment ALPS (``Any Light Particle Search experiment) at DESY \cite{Redondo:2010dp} ($\omega=2.33$ eV, $B_{ext}=5$ T and $L=4.21$ m), the probability of shining a photon through a wall via an intermediate axion with $m_{a}\sim10^{-5}$ eV is about $P_{\gamma\rightarrow a\rightarrow\gamma}\simeq10^{-53}$ and the coupling to photons is $g_{a\gamma\gamma}\sim10^{-15}\,\mathrm{GeV^{-1}}$. The upgrade of LSTW experiments intends to achieve a sensitivity in
the photon-WISP coupling of $g_{a\gamma\gamma}<10^{-11}\,\mathrm{GeV^{-1}}$, corresponding to an improvement of 3 orders of magnitude comparing with ALPS 2010 results \cite{Ehret:2010ki,Ehret:2010mh}. 

In our case, the Higgs coupling to two photons is expressed by the following term in the Lagrangian \cite{Melnikov:1993tj}:
\begin{equation}
\mathcal{L}_{h\gamma\gamma}=\frac{g_{h\gamma\gamma}}{4}\,h\,F_{\mu\nu}F^{\mu\nu}~,\label{higgs photon photon}
\end{equation}
where the coupling $g_{h\gamma\gamma}$ has the form
\begin{equation}
g_{h\gamma\gamma}=\frac{\alpha_{QED}\,F}{\pi\,\mathrm{v}}~,\label{coupling higgs photon}
\end{equation}
with $F$ given by Eq. (\ref{F factor}). Using the mixing $\phi-h$ in Eq. (\ref{mixing h and phi}) we can write
\begin{equation}
\mathcal{L}_{\phi\gamma\gamma}=-\frac{g_{h\gamma\gamma}}{4}\,\epsilon\,\tilde{\phi}\,F_{\mu\nu}F^{\mu\nu}~,\label{phi photon photon}
\end{equation}
and we may define the effective coupling of the dark scalar field to photons, $g_{\phi\gamma\gamma}$, as:
\begin{equation}
g_{\phi\gamma\gamma}=g_{h\gamma\gamma}\,\epsilon~,\label{coupling phi photon}
\end{equation}
where $\epsilon=\epsilon_{-}$ for the negative coupling between $\phi$ and $h$, and $\epsilon=\epsilon_{+}$ otherwise. For the former, we find
\begin{equation}
g_{\phi\gamma\gamma}^{\left(-\right)}\simeq2\times10^{-26}\,\left(\frac{m_{\phi}}{10^{-5}\,\mathrm{eV}}\right)\left(\frac{x_{DM}}{0.5}\right)^{-1}\,\mathrm{GeV^{-1}}~,\label{coupling SSB}
\end{equation}
while for the case without spontaneous symmetry breaking we obtain:
\begin{equation}
g_{\phi\gamma\gamma}^{\left(+\right)}\simeq8\times10^{-49}\,\left(\frac{m_{\phi}}{10^{-5}\,\mathrm{eV}}\right)\,\mathrm{GeV^{-1}}~,\label{coupling NSSB}
\end{equation}
where we have used the cosmological value of the dark matter energy density. Hence, even for the most promising case with spontaneous symmetry breaking the effective coupling to photons of our dark scalar is 11 orders of magnitude below the corresponding axion-photon coupling for the same mass, with the probability for LSTW thus being suppressed by $\sim 10^{-44}$ with respect to the axion. This makes our dark scalar much harder to detect in LSTW experiments, and hence requiring a very substantial improvement in technology. 


\subsubsection{Oscillation of fundamental constants}

A light dark scalar, $10^{-22}\,\mathrm{eV}<m_{\phi}<0.1\,\mathrm{eV}$, behaves as a coherently oscillating field on galactic scales, which may cause the variation of fundamental constants, namely the electron's mass, $m_{e}$, and the fine structure constant, $\alpha_{QED}$.

The Standard Model Yukawa interactions can be generically written as:
\begin{equation}
\mathcal{L}_{hff}=\frac{g_{f}}{\sqrt{2}}\,\bar{f}\,f\,h~,\label{higgs fermion fermion}
\end{equation}
where $g_{f}$ is the Yukawa coupling. Using the $\phi-h$ mass mixing in Eq. (\ref{mixing h and phi}), the electron's mass after the EWPT is thus given by:
\begin{equation}
m_{e}\simeq\frac{g_{f}\,\mathrm{v}}{\sqrt{2}}\,\left(1-\frac{d_{m_{e}}}{\sqrt{2}\,M_{Pl}}\,\phi_{CDM}\right)~,\label{mass electron}
\end{equation}
where $\phi_{CDM}$ corresponds to the present amplitude of the scalar field  and $d_{m_{e}}$ is a dimensionless quantity that works as an effective ``dilatonic'' coupling
and is normalized by the Planck mass, such that \cite{Damour:2010rp,VanTilburg:2015oza}:
\begin{equation}
\frac{d_{m_{e}}}{\sqrt{2}\,M_{Pl}}=\frac{\epsilon_{\pm}}{\mathrm{v}}.\label{de}
\end{equation}
In the case where $\phi$ undergoes spontaneous symmetry breaking, the dilatonic coupling $d_{m_{e}}^{\left(-\right)}$ is  
\begin{equation}
d_{m_{e}}^{\left(-\right)}\simeq6\times10^{-16}\,\left(\frac{m_{\phi}}{10^{-15}\,\mathrm{eV}}\right)\,\left(\frac{x_{DM}}{0.5}\right)^{-1}~,\label{mass e var B}
\end{equation}
while, for a positive Higgs-portal coupling, 
\begin{equation}
d_{m_{e}}^{\left(+\right)}\simeq2\times10^{-38}\,\left(\frac{m_{\phi}}{10^{-15}\,\mathrm{eV}}\right)~.\label{mass e var NB}
\end{equation}
Similarly, we may compute the variation of the fine structure constant due to the oscillating dark scalar field. We have seen that the interaction with the Higgs
boson introduces a small coupling to the electromagnetic field (Eq.(\ref{phi photon photon})), which will induce a variation on the fine structure constant of the
form \cite{Stadnik:2016zkf}:
\begin{equation}
\alpha'_{QED}\simeq\alpha_{QED}\left(1+\frac{d_{\alpha}}{\sqrt{2}\,M_{Pl}}\,\phi_{CDM}\right)~,\label{alpha QED var}
\end{equation}
where, after the EWPT, 
\begin{equation}
d_{\alpha}^{\left(-\right)}\simeq7\times10^{-16}\,\left(\frac{m_{\phi}}{10^{-15}\,\mathrm{eV}}\right)\,\left(\frac{x_{DM}}{0.5}\right)^{-1}~,\label{variation fine struct B}
\end{equation}
and 
\begin{equation}
d_{\alpha}^{\left(+\right)}\simeq3\times10^{-38}\,\left(\frac{m_{\phi}}{10^{-15}\,\mathrm{eV}}\right)~.\label{variation fine struct NB}
\end{equation}
The dilatonic couplings $d_{m_e}$ and $d_{\alpha}$ have thus comparable values, both in the case with and without spontaneous symmetry breaking, which is a distinctive prediction of our model.

Despite the smallness of these dilatonic couplings, there are a few ongoing experiments designed to detect oscillations of fundamental couplings, using, for instance, atomic clock
spectroscopy measurements and resonant-mass detectors \cite{VanTilburg:2015oza,Stadnik:2014tta,Arvanitaki:2014faa,Stadnik:2015kia,Stadnik:2015xbn,Stadnik:2016zkf,Hees:2016gop,Arvanitaki:2015iga,Arvanitaki:2016fyj}.  For instance, atomic spectroscopy using dysprosium can probe the $d_{m_e}$ coupling in the mass range $10^{-24}\,\mathrm{eV}\lesssim m_{\phi}\lesssim10^{-15}\,\mathrm{eV}$, for $10^{-7}\lesssim d_{m_e}\lesssim10^{-1}$. In turn, the current AURIGA experiment may reach $10^{-5}\lesssim d_{m_e},\,d_{\alpha}\lesssim10^{-3}$ in the narrow interval $10^{-12}\,\mathrm{eV}\lesssim m_{\phi}\lesssim10^{-11}\,\mathrm{eV}$, whereas the planned DUAL detector intends to achieve, for $10^{-12}\,\mathrm{eV}\lesssim m_{\phi}\lesssim10^{-11}\,\mathrm{eV}$, sensitivity to detect $10^{-6}\lesssim d_{m_e},\,d_{\alpha}\lesssim10^{-2}$. From Eqs. (\ref{mass e var B})-(\ref{variation fine struct NB}), we conclude that current technology is not enough to probe our model, which may nevertheless serve as a motivation to significantly improve the sensitivity of such experiments.

We should also point out that ultra-light scalars with mass below $10^{-10}$ eV can lead to superradiant instabilities in astrophysical black holes that may lead to distinctive observational signatures, such as gaps in the mass-spin Regge plot as first noted in \cite{Arvanitaki:2009fg}. However, these instabilities should only be able to distinguish our Higgs-portal dark matter candidate from other ultra-light scalars, such as axions and axion-like particles, if non-gravitational interactions also play a significant dynamical role (see e.g. Ref. \cite{Rosa:2017ury}).


\section{Cosmological implications of the spontaneous symmetry breaking} \label{Cosmo implications}

Throughout our analysis, we have assumed that the model exhibits a U(1) symmetry which,
in the case of a negative sign for the coupling between the Higgs and the dark scalar,
is spontaneously broken at the EWPT. The implications of such symmetry breaking
depend on whether it is a global or a local symmetry. The Lagrangian density
is of the form:
\begin{equation}
\mathcal{L=}\partial_{\mu}\Phi\partial^{\mu}\Phi^{\dagger}-\mathcal{L}_{int}~,\label{lagrangian total}
\end{equation}
where $\mathcal{L}_{int}$ is given by Eq. (\ref{Lagrangian}).

First, let us suppose that the Lagrangian density is invariant under
a global U(1) symmetry, i.e.~that the complex dark scalar is
invariant under the following transformation:
\begin{equation}
\Phi\rightarrow e^{i\alpha}\Phi~,\label{global u1 symmetry}
\end{equation}
where $\alpha$ is a constant parameter. Expanding the kinetic term for $\Phi=\phi e^{i\theta}/\sqrt{2}$ and introducing
the rescaled field $\sigma=\phi_{0}\,\theta$, we find the following kinetic and interaction terms between the dark scalar, $\phi$, and the associated Goldstone boson, $\sigma$:
\begin{equation}
\partial_{\mu}\Phi\partial^{\mu}\Phi^{\dagger}=\frac{1}{2}\left[\partial_{\mu}\phi\partial^{\mu}\phi+\frac{\phi^{2}}{\phi_{0}^{2}}\partial_{\mu}\sigma\partial^{\mu}\sigma+2\,\frac{\phi}{\phi_{0}}\,\partial_{\mu}\sigma\partial^{\mu}\sigma+\partial_{\mu}\sigma\partial^{\mu}\sigma\right]~.\label{kinetic term sigma}
\end{equation}
The third term on the right-hand-side of Eq.~(\ref{kinetic term sigma}) allows for the decay
of $\phi$ into two massless $\sigma$ particles, which could imply the complete evaporation of the dark scalar condensate. The corresponding decay width is then:
\begin{equation}
\Gamma_{\phi\rightarrow\sigma\sigma}=\frac{1}{64\,\pi}\,\frac{m_{\phi}^{3}}{\phi_{0}^{2}}~,\label{decay width Goldstone}
\end{equation}
such that, using the relation between $g$ and $\lambda_{\phi}$ in Eq. (\ref{g lambda neg}),
\begin{align}
\Gamma_{\phi\rightarrow\sigma\sigma} & \simeq7\times10^{21}\,\left(\frac{x_{DM}}{0.5}\right)^{-1/2}\,\lambda_{\phi}^{5/4}\,\mathrm{sec^{-1}}~,\label{decay width simpl}
\end{align}
corresponding to 
\begin{equation}
\tau_{\phi\rightarrow\sigma\sigma}\simeq10^{-22}\,\left(\frac{x_{DM}}{0.5}\right)^{1/2}\,\lambda_{\phi}^{-5/4}\,\mathrm{sec}~.\label{decay time}
\end{equation}
The field is sufficiently long-lived to account for dark matter if $\tau_{\phi\rightarrow\sigma\sigma}>t_{\mathrm{univ}}\sim4\times10^{17}\,\mathrm{sec}$, placing an upper bound on $\lambda_{\phi}$: 
\begin{equation}
\lambda_{\phi}<2\times10^{-32}\,\left(\frac{x_{DM}}{0.5}\right)^{2/5}~,\label{constraint on lambda}
\end{equation}
and thus restricting the viable range for the dark matter mass to:
\begin{equation}
m_{\phi}\lesssim5\,\mathrm{eV}.\label{DM new constraint}
\end{equation}
Therefore, if the U(1) symmetry were global, there would be a stricter constraint
on the value of $\lambda_{\phi}$, and our dark matter
model could not, in particular, explain the 3.5 keV line. 

However, we may consider a local U(1) gauge
symmetry, where the Lagrangian is invariant under 
\begin{equation}
\Phi\rightarrow e^{i\alpha\left(x\right)}\Phi~,\label{local u1 symm}
\end{equation}
which can be achieved by introducing a gauge field and covariant derivative such that:
\begin{equation}
 D_{\mu}\Phi=\partial_{\mu}\Phi-i\,e'\,A_{\mu}\Phi,\label{covariant derivative}
\end{equation}
\begin{equation}
A_{\mu}\rightarrow A_{\mu}+\frac{\partial_{\mu}\alpha\left(x\right)}{e'}~,\label{gauge transform}
\end{equation}
where $e'$ denotes the associated gauge coupling. In this case, expanding the scalar kinetic term in the unitary gauge, where the Goldstone boson is manifestly absorbed into the longitudinal component of the ``dark photon" gauge field, we get:
\begin{align}
D_{\mu}\Phi D^{\mu}\Phi^{\dagger} & =\frac{1}{2}\partial_{\mu}\phi\partial^{\mu}\phi+\frac{1}{2}\,e'^{2}\,A_{\mu}A^{\mu}\,\phi^{2}+\frac{1}{2}\,e'^{2}\,\phi_{0}\,\phi\,A_{\mu}A^{\mu}+\frac{1}{2}\,e'^{2}\,\phi_{0}^{2}\,A_{\mu}A^{\mu},\label{kinetic term local}
\end{align}
where the second and the third terms correspond to dark scalar-dark photon interactions and the last term gives the mass of the dark photon, $\gamma'$:
\begin{equation}
m_{\gamma'}=e'\,\phi_{0}~.\label{massive photon}
\end{equation}
For $m_{\gamma'}<m_{\phi}/2$, the dark scalar may decay into two dark photons with a decay width
\begin{align}
\Gamma_{\phi\rightarrow\gamma'\gamma'} \simeq\frac{1}{16\pi}\,\sqrt{1-\frac{4\,m_{\gamma'}^{2}}{m_{\phi}^{2}}}\,\frac{m_{\gamma'}^{4}}{m_{\phi}\,\phi_{0}^{2}}\,\left(3+\frac{1}{4}\,\frac{m_{\phi}^{4}}{m_{\gamma'}^{4}}-\frac{m_{\phi}^{2}}{m_{\gamma'}^{2}}\right)~.\label{decay gauge bosons}
\end{align}
This remains finite even in the limit $e'\rightarrow 0$ and, in fact, it tends to the value in Eq. (\ref{decay width Goldstone}), since in this limit the symmetry is global and the decay into massless Goldstone bosons is allowed.

On the other hand, for $m_{\gamma'}>m_{\phi}/2$, the dark scalar's decay into dark photon pairs becomes kinematically forbidden. This requires:
\begin{equation}
e'>\sqrt{2\,\lambda_{\phi}}~,\label{constraint on gy}
\end{equation}
which does not pose a significant constraint, given the magnitude of the scalar self-couplings considered in our analysis. Note that even if this condition is satisfied the scalar self-coupling is stable against gauge radiative corrections, since $\delta\lambda_\phi/\lambda_\phi \sim {e'^4}/\lambda_\phi \gtrsim \lambda_\phi $ and the self-coupling is typically very small in the scenarios under consideration.

It is important to mention that, similarly to the case studied in section \ref{perturbative prod}, before
the EWPT, the dark scalar's oscillations may also lead to the production of dark photons, through the second term in Eq. (\ref{kinetic term local}). This leads to an upper bound on the squared gauge coupling comparable to that obtained for the scalar self-interactions in Eq. (\ref{upper bound on lambda}), and thus nevertheless compatible with the stability condition Eq.~(\ref{constraint on gy}).

Notice that the spontaneous breaking of a U(1) symmetry can lead to the generation of cosmic strings at the EWPT. The ratio between the energy density of such cosmic strings, $\rho_{s}$, and the background
density, $\rho_{c}$, reads \cite{Kolb:1990vq}
\begin{equation}
\frac{\rho_{s}}{\rho_{c}}\sim G\mu\simeq10^{-6}\left(\frac{\phi_{0}}{10^{16}\,\mathrm{GeV}}\right)^{2},\label{cosmic strings ratio}
\end{equation}
where $G$ is Newton's gravitational constant and $\mu$ is the
string's energy per unit length. Nevertheless, as we reported in Ref. \cite{Cosme:2017cxk},
$\phi_{0}\ll10^{16}\,\mathrm{GeV}$ even for very suppressed $\lambda_{\phi}$
(for instance, $\phi_{0}$ is always smaller than $10^{16}\,\mathrm{GeV}$
for $\lambda_{\phi}>10^{-66}$), implying that the ratio Eq. (\ref{cosmic strings ratio})
is negligible and that, therefore, this does not pose additional constraints on the model.

Finally, we note that all the dynamics and predictions of our model could be achieved, however, by considering only a real scalar field with a $\mathbb{Z}_{2}$ symmetry. Spontaneous breaking of such a symmetry then leads to the generation of domain walls at the EWPT, which could have disastrous consequences for the cosmological dynamics. However, it has been argued that a statistical bias in the initial configuration of the scalar field could effectively yield a preferred minimum and thus make the domain wall network decay  \cite{Larsson:1996sp}. In particular, according to Ref. \cite{Larsson:1996sp}, inflation itself may produce such a bias through the quantum fluctuations of the scalar field that become frozen on super-horizon scales. Since our dark scalar never thermalizes with the ambient cosmic plasma, such a bias could survive until the EWPT and therefore lead to the destruction of any domain wall network generated during the phase transition. A detailed study of the evolution of domain wall networks in the context of the proposed scenario is beyond the scope of this work, but this nevertheless suggests that a real scalar field, with no additional undesired degrees of freedom, may yield a consistent cosmological scenario for dark matter with spontaneous symmetry breaking.

In summary, we conclude that the cosmological consistency of the scenarios with spontaneous symmetry breaking requires either a complex scalar field transforming under a gauged U(1) symmetry with sufficiently large gauge coupling or possibly a real scalar field transforming under a $\mathbb{Z}_{2}$ symmetry.


\section{Conclusions}

In this work we have analyzed the dynamics and the phenomenology of a non-thermal dark matter candidate corresponding to an oscillating scalar field that, as all other known elementary particles, acquires mass solely through the Higgs mechanism. The model assumes an underlying scale invariance of the interactions that is broken by an unspecified mechanism to yield the electroweak and Planck scales. 

The dynamics of the scalar field may be summarized in the following way. During inflation, the field acquires a Hubble-scale mass through a non-minimal coupling to gravity that drives the classical field towards the origin, while de Sitter quantum fluctuations generate a sub-Hubble field value on average. The Hubble scale mass also suppresses potentially significant cold dark matter isocurvature modes in the CMB anisotropies spectrum. After inflation, this non-minimal coupling plays no significant role in the dynamics, which is driven essentially by the scalar potential. After inflation, the field oscillates in an quartic potential, behaving as dark radiation, until the electroweak phase transition. At this point the field acquires mass through the Higgs mechanism, and starts behaving as non-relativistic matter. 

If the Higgs portal coupling is positive, the dark scalar oscillates about the origin in a quadratic potential until the present day, while for a negative coupling it undergoes spontaneous symmetry breaking and oscillates about a non-vanishing vacuum expectation value. Whereas in the former case the present dark matter abundance depends on all model parameters, including the non-minimal coupling to gravity and the scale of inflation, in the latter scenario it is determined uniquely the the Higgs-portal coupling and the dark scalar's self-interactions. The suppression of particle production processes that could lead to the oscillating condensate's evaporation places strong constraints on the value of these couplings, and we generically find that the dark scalar's mass must lie below the MeV scale. It should be pointed out that the aim of this paper is not to explain the smallness of the dark matter couplings. In fact, in the parametric regime that we are interested in ($g/\lambda^{1/4}\sim10^{-3}-10^{-2}$), both $g$ and $\lambda_{\phi}$ are technically natural, since their relation is not significantly affected by radiative corrections, which makes them as natural as the electron Yukawa coupling. Nevertheless, small couplings can be naturally achieved in theories with extra dimensions, as we have shown in Ref. \cite{Bertolami:2016ywc}.  

While there are several phenomenological consequences of the Higgs-portal interactions of the dark scalar that could allow for its detection, the required suppression of the latter makes this rather challenging in practice. Possible laboratory signatures include invisible Higgs decays, dark scalar-photon oscillations and induced oscillations of the fine structure constant and the electron mass. Indirect astrophysical signatures are also possible, namely dark matter annihilation or decay into photons. All these processes could lead to a robust identification of our proposed dark matter candidate, but unfortunately they lie generically below the reach of current technology, even for the case with spontaneous breaking that is generically easier to detect experimentally. An interesting exception that we have already reported in Ref. \cite{Cosme:2017cxk} is the decay of a 7 keV dark scalar in the case with spontaneous symmetry breaking, whose decay into photons may naturally explain 3.5 keV emission line observed in the galactic centre, the Andromeda galaxy and the Perseus cluster.   

In summary, the proposed oscillating scalar field is a viable dark matter candidate with distinctive observational and experimental signatures, constituting a promising alternative to WIMPs, which have so far evaded detection.  While it is certainly amongst the ``darkest" dark matter candidates available in the literature, there may already be astrophysical hints for its existence, and we hope that this work may motivate future technological developments that may allow for testing its implications in the laboratory.

\label{conclusions}


\section*{Acknowledgements}
C.\,C. is supported by the Funda\c{c}\~{a}o para a Ci\^{e}ncia e Tecnologia (FCT) grant PD/BD/114453/2016. J.\,G.\,R. is supported by the FCT Investigator Grant No.~IF/01597/2015 and partially by the H2020-MSCA-RISE-2015 Grant No. StronGrHEP-690904 and by the CIDMA Project No.~UID/MAT/04106/2013. We would like to thank Igor Ivanov and his colleagues from the CFTP, IST-Lisbon for interesting discussions.


\appendix

\section{Effects of the non-minimal coupling to gravity on an oscillating scalar field \label{sec:Continuity-equation}}

Consider a homogeneous scalar field $\phi$ that is oscillating about the minimum of a potential $V(\phi)$ much faster than Hubble expansion, i.e.~in the regime where the effective field mass $m_\phi \gg H$. In this regime we then have that:
\begin{eqnarray}
{d\over dt }\langle \phi\dot\phi \rangle = \langle \dot\phi^2\rangle + \langle \phi \ddot\phi\rangle =  \langle \dot\phi^2\rangle  - \langle V'(\phi )\phi\rangle = 0~,
\end{eqnarray}
in a stationary configuration for quantities averaged over the oscillating period, where we have used the equation of motion $\ddot \phi = -V'(\phi)$ discarding the sub-leading effects of expansion. We thus obtain the virial theorem:
\begin{align}
\langle \dot{\phi}^{2}\rangle  & =\langle \phi V'\left(\phi\right)\rangle \nonumber \\
 & =n\langle V\left(\phi\right)\rangle,\label{virial theorem}
\end{align}
where the second line is valid for a potential of the form $V\left(\phi\right)\sim\phi^{n}$. Since $R \sim H^2 \ll m_\phi^2$ and $\dot\phi \sim m_\phi \phi$, we may discard the terms  that include $R$ or $H$ explicitly in the expressions for the field's energy density and pressure given in Eq. (\ref{energy density}), assuming $\xi\lesssim 1$. We may then write these quantities approximately as:

\begin{eqnarray}
\rho_{\phi}&\simeq&\frac{\dot{\phi}^{2}}{2}+V\left(\phi\right)~,\nonumber\\
p_{\phi} & \simeq&\frac{1}{2}\,\dot{\phi}^{2}-V\left(\phi\right)-4\xi\left(\dot{\phi}^{2}-\phi V'\left(\phi\right)\right)~.\label{pressure approx}
\end{eqnarray}
It is easy check, using the virial relation (\ref{virial theorem}), that the term proportional to $\xi$ vanishes on average, thus yielding the usual expressions for the energy density and pressure of a homogeneous scalar field in minimally-coupled general relativity. This leads to the following equation of state parameter $w$:
\begin{align}
w & \equiv\frac{p}{\rho}\nonumber \\
 & \simeq\frac{n-2}{n+2}.\label{w}
\end{align}
Also, for $m_\phi \gg H$, the field's equation of motion reads
\begin{equation}
\ddot{\phi}+3H\dot{\phi}+V'\left(\phi\right)=0~,\label{eom without R}
\end{equation}
and, multiplying both sides by $\dot{\phi}$, we obtain the standard continuity equation:
\begin{equation}
{d\over dt}\left({1\over 2}\dot\phi^2+V(\phi)\right)+3H\dot{\phi}^{2}=\dot{\rho_\phi}+3H(\rho_\phi+p_\phi)=0~.\label{continuity eq}
\end{equation}
Hence, the non-minimal coupling to gravity $\xi$ does not affect the field's dynamics and properties in an underdamped regime.




\begin{thebibliography}{9}
\bibitem{foot1}
Since since $\phi\ll M_{Pl}$ and $\xi\lesssim1$, $f\left(\phi\right)\simeq1$ and there are no significant differences between the Jordan and Einstein frames. In addition, $\phi$ is not the inflaton and, therefore, the gravitational dynamics during inflation is not altered by $\phi$, regardless its non-minimal coupling 
to curvature.
\end{thebibliography}
\end{document}